\documentclass[%
 aip,
 amsmath,amssymb,
 reprint,%
]{revtex4-1}

\usepackage{graphicx}
\usepackage{dcolumn}
\usepackage{bm}

\usepackage[utf8]{inputenc}
\usepackage[T1]{fontenc}
\usepackage{mathptmx}
\usepackage{hyperref}
\usepackage{amsmath} 
\usepackage{multirow}
\usepackage{color}
\usepackage{ulem}
\newcommand{\angstrom}{\text{\normalfont\AA}}
\begin{document}

\preprint{AIP/123-QED}

\title[\small Ring Polymer Molecular Dynamics and Active Learning of Moment Tensor Potential for Gas-Phase Barrierless Reactions: Application to S + H$_2$]{Ring Polymer Molecular Dynamics and Active Learning of Moment Tensor Potential for Gas-Phase Barrierless Reactions: Application to S + H$_2$}

\author{Ivan S. Novikov}
 \email{i.novikov@skoltech.ru}
\author{Alexander V. Shapeev}%
 \email{a.shapeev@skoltech.ru}
\affiliation{ 
Skolkovo Institute of Science and Technology, Skolkovo Innovation Center, Nobel St. 3, Moscow 143026, Russia
}%

\author{Yury V. Suleimanov}
 \email{ysuleymanov@cyi.ac.cy}
 \altaffiliation[Also at ]{Department of Chemical Engineering, Massachusetts Institute of Technology, Cambridge, Massachusetts 02139, United States}
\affiliation{%
Computation-based Science and Technology Research Center, Cyprus Institute, 20 Kavafi Street, Nicosia 2121, Cyprus
}%

\date{\today}

\begin{abstract}
Ring polymer molecular dynamics (RPMD) has proven to be an accurate approach for calculating thermal rate coefficients of various chemical reactions. For wider application of this methodology, efficient ways to generate the underlying full-dimensional potential energy surfaces (PESs) and the corresponding energy gradients are required. Recently, we have proposed a fully automated procedure based on combining the original RPMDrate code with active learning for PES on-the-fly using moment tensor potential and successfully applied it to two representative thermally activated chemical reactions [I. S. Novikov, Y. V.  Suleimanov, A. V. Shapeev, Phys. Chem. Chem. Phys. {\bf 20}, 29503-29512 (2018)]. In this work, using a prototype insertion chemical reaction S + H$_2$, we show that this procedure works equally well for another class of chemical reactions. We find that the corresponding PES can be generated by fitting to less than 1500 automatically generated structures while the RPMD rate coefficients show deviation from the reference values within the typical convergence error of RPMDrate. We note that more structures are accumulated during the  real-time  propagation  of  the dynamic factor (the recrossing factor) as opposed to the previous study. We also observe that relatively flat free energy profile of the along the reaction coordinate before entering the complex-formation well can cause issues with locating the maximum of the free energy surface for less converged PESs.  However,  the final RPMD rate coefficient is independent of the position of the dividing surface that makes it invulnerable to this problem, keeping the total number of necessary structures within a few thousand. Our work concludes that, in future, the proposed methodology can be applied to realistic complex chemical reactions with various energy profiles. 

\end{abstract}

\maketitle

\section{Introduction}
Thermal rate coefficients for elementary chemical reactions are the key input parameters in chemical kinetics models used to simulate various fundamental and applied processes relevant to astrochemistry, atmospheric and combustion chemistry, pyrolysis etc. Their experimental measurements can be confronted with certain difficulties, such as, {\it e.g.}, low temperatures or inability to isolate/stabilize products or even reactants. Moreover, chemical kinetic models contain thousands of chemical reactions~\cite{kida}, experimental validation of each of them is an extremely onerous task. 

Recent progress in electronic structure and rate theories suggests that computer simulations are becoming an inexpensive alternative to experiment. 
Among dynamics approaches, ring polymer molecular dynamics (RPMD) stands out sharply against the background due to its consistent and reliable performance across all the chemical systems studied so far~\cite{rpmdreview}. The RPMD method is based on an {\it ad hoc} idea~\cite{craig2004} of approximating quantum real-time Kubo-transformed correlation functions used to describe various dynamical processes, such as chemical reactions~\cite{craig2005_1,craig2005_2} by classical ones originating from the isomorphism between the quantum statistical mechanics of a quantum system and the classical statistical mechanics of a fictitious ring polymer.  The ring polymer is composed of $n_{\rm beads}$ classical copies of the original system (beads) connected by harmonic springs. Hence, RPMD is a purely classical molecular dynamics but in an extended $n_{\rm beads}$ imaginary time path integral phase space \cite{habershon2013}. Despite its {\it ad hoc} nature, 
RPMD provides exact solutions in certain limits~\cite{craig2004,braams06,habershon2013} and immediately found its application in simulations of condensed phase systems~\cite{habershon2013}. Later on, it was demonstrated that RPMD offers a very reliable and accurate way to calculate thermal rate coefficients for various bimolecular chemical reactions in wide temperature ranges. This includes both thermally activated chemical reactions (with energy barrier along the reaction path, such as prototype atom-diatom~\cite{collepardo2009,collepardo2009e,perezdetudela2012,suleimanov2013a,perezdetudela2014a} and more complex systems~\cite{suleimanov2011,allen2013,li2013a,li2013b,li2013c,espinosa2014,dong2015,espinosa2015,arseneau2016,zhang2016,xie2016,espinosa2017}) and chemical reactions with deep wells due to complex-formation (such as typical insertion triatomics~\cite{rpmd_ins1,rpmd_ins2,rpmd_ins3,rpmd_ins4,rpmd_ins5,rpmd_ins6,rpmd_ins7} as well as polyatomic complex-forming systems ~\cite{espinosa2013,perezdetudela2014b,octavio2018,sa2018,rpmd_ins2019a,rpmd_ins2019b}).  

RPMD is a full dimensional approach based on running trajectories on the underlying global potential energy surfaces (PESs). This is, on the one hand, an advantage of RPMD as it is able to capture automatically various features along the reaction path such as deep tunneling,~\cite{perezdetudela2014a} complex zero point energy effects,~\cite{perezdetudela2012,espinosa2014} role of asymptotic interactions at low temperatures~\cite{rpmd_ins7}, etc. On the other hand, the requirement of availability of a global PES constitutes its limitation as only a very few systems have pre-constructed PESs available~\cite{rpmdreview}. For the RPMD rate theory to become widely used, efficient ways to couple RPMD with electronic structure evaluations are therefore required.  Recently, we proposed  a methodology for fully automated calculation of thermal rate coefficients for gas phase chemical reactions which
is based on combining RPMD with the machine-learning interatomic potentials (namely, moment tensor potentials, MTPs) actively learning (AL) on-the-fly (AL-MTP) ~\cite{novikov2018}. Initially, MTPs were proposed for single-component systems~\cite{Shapeev2016-MTP} and then generalized to the case of multi-component systems~\cite{Gubaev2018-MTPR}. An AL algorithm allows one to construct a training set needed for training a machine-learning interatomic potentials automatically, without the need in manual parametrization of potentials based on many iterations of trial and error. Popular existing active learning methods use query by committee approaches~\cite{artrith2012high,zhang2019-e-active,smith2018-active-learning} and Bayesian predictive variance~\cite{jinnouchi2019-perovskites}. In this paper we will rely on the D-optimality-based AL algorithm \cite{Podryabinkin2017-AL}. The AL-MTP method was also successfully applied for solving various multiscale condensed phase problems, such as diffusion of point defects in materials~\cite{novoselov2019-Diffusion}, crystal structure prediction~\cite{podryabinkin2019-crystals}, prediction of new stable alloys~\cite{gubaev2019-alloys}, and the study of the phase transitions of the high-entropy alloy~\cite{kostiuchenko2019-HEA}. We refer to the combination of RPMD and AL-MTP methods as RPMD-AL-MTP. For two representative thermally activated chemical reactions (OH + H$_2$ and CN + CH$_4$), RPMD-AL-MTP displayed a remarkable accuracy and agreement with the previous RPMD results \cite{novikov2018} that encourages its future application. 

Following our previous study of thermally activated reactions, we extend it in the present work to one of the prototypical insertion reactions,~\cite{rpmd_ins1,rpmd_ins2}, which proceed through deep complex formation well, namely, X + H$_2$ $\rightarrow $ HX + H, where in the present work X = S($^1$D). We show that the previously proposed computational strategy works equally well and maintains the accuracy for calculating thermal rate coefficients for this class of reactions.

\section{General methodology}

\subsection{Ring Polymer Molecular Dynamics}

A detailed description of the RPMD rate theory can be found in Refs. ~\cite{suleimanov2011,suleimanov2013b} and its practical implementation for various benchmark systems is summarized in the recent review ~\cite{rpmdreview}. Technical aspects of the computational procedure developed for calculating thermal rate coefficients of any bimolecular chemical reaction is well-documented in the manual of general RPMDrate code developed by one of us (Y.V.S.)~\cite{suleimanov2013b} . 

In brief, the ring polymer Hamiltonian of a system consisting of $N$ atoms with fictitious ring polymers of $n_{\rm beads}$  is written in atomic cartesian coordinates as (in atomic units)
\begin{eqnarray}
H({\bf p},{\bf q}) & = & \sum_{i=1}^N\sum_{j=1}^{n_{\rm beads}}\left( \frac{{p_i^{(j)}}^2}{2m_i} + \frac{1}{2}m_i\omega_n^2\left | q_i^{(j)} - q_i^{(j-1)}\right |^2 \right )   \nonumber \\
& & + \sum_{j=1}^{n_{\rm beads}}V(q_1^{(j)},q_2^{(j)},...,q_N^{(j)}), 
\label{eqn1}
\end{eqnarray}
with $q_i^{(j)}$ and $p_i^{(j)}$ being the position and momentum of the $j$-th bead of the $i$-th atom of the system, correspondingly, and $q_i^{(0)}\equiv q_i^{(n_{\rm beads})}$ ensures that the polymer is closed. The force constant of the harmonic springs is $\omega_n={\beta\hbar}/{n_{\rm beads}}$ and $\beta=1/k_BT$, where $T$ is the temperature of the system. 

We introduce a dividing surface $s(\bf{q})=0$ to separate reactants and products, such that the latter is in the  $s>0$ region, and the reaction coordinate $\bar{s}(\bf{q})=s(\bar{q}_1,...,\bar{q}_N)$ is defined using the centroid variables $\bar{q}_i=\frac{1}{n_{\rm beads}}\sum_{j=1}^{n_{\rm beads}} q_i^{(j)}$. As explained in Suleimanov {\it et al.}~\cite{suleimanov2013b,rpmdreview}, the method uses a formalism based on two dividing surfaces $s_1$ (in the reaction active region) and $s_0$ (in the reactants asymptote). 

The correlation function formalism used in the computational procedure for the RPMD rate coefficient calculation is based on the $t\rightarrow +\infty$ limit of the ring polymer flux-side correlation function $c_{\rm fs}$~\cite{suleimanov2011}. The rate coefficient is then expressed using the Bennett-Chandler factorization~\cite{bennett77,chandler78} as 
\begin{eqnarray}
k_{\rm RPMD}(T) = \kappa(s_1)k_{\operatorname{cd-TST}}(s_1)=\kappa(s_1)p(s_1,s_0)k_{\operatorname{cd-TST}}(s_0).
\label{eqn2}
\end{eqnarray}
The dividing surface $s_1$ is situated near the free energy maximum, its general expressions can be found in Ref.~\cite{suleimanov2013b}. The second dividing surface, $s_0$, is localized in the asymptotic reactant valley and is  defined as  $s_0(\bar{\bf{q}}) = R_\infty - | \bar{\bf{R}} | = 0$, $\bar{\bf{R}}$ being the centroid of the Jacobi vector that connects the center of mass of the two reactants and $R_\infty$ is an asymptotic distance large enough to make interaction between them negligible.

The first factor in Eq.~\ref{eqn2} is an $n_{\rm beads}$ ring polymer transmission coefficient for a dividing surface $s_1$
\begin{eqnarray}\label{eq:rpmdtc}
\kappa(s_1) = \frac{c_{\rm fs}(t\rightarrow\infty;s_1)}{c_{\rm fs}(t\rightarrow0_+;s_1)}.
\label{eqn3}
\end{eqnarray}

The second factor is the ratio of two short-time limits of ring polymer flux-side correlation functions for different dividing surfaces which can also be expressed in terms of the centroid potential of mean force (PMF), or free energy, $W(s)$~\cite{suleimanov2011},
\begin{eqnarray}
p(s_1,s_0) 	\equiv \frac{c_{\rm fs}(t\rightarrow0_+;s_1)}{c_{\rm fs}(t\rightarrow0_+;s_0)}=e^{-\beta[W(s_1) - W(s_0)]}.
\end{eqnarray}

The third term is the centroid density transition state theory (cd-TST)~\cite{gillan87a,gillan87b,voth89} rate coefficient for the dividing surface $s_0$ which is expressed analytically as
\begin{eqnarray}
k_{\operatorname{cd-TST}}(s_0) &=& 4\pi R_\infty^2\left (\frac{1}{2\pi\beta\mu_R}\right )^{1/2},
\label{eqn4}
\end{eqnarray}
where $\mu_R$ is the reduced mass of the reactants. Thus, after calculating the three terms, we can calculate the RPMD rate coefficient $k_{\rm RPMD}$.
In practice, the first two factors ($\kappa $ and $p$) are calculated at the maximum free energy W($\xi ^{\ddagger}$) value along 
the reaction coordinate $\xi $ which is an interpolating function used to connect the two dividing surfaces, $\xi (q)$ = $s_0(q)/(s_0(q) - s_1(q))$ and varies from $\xi  \rightarrow 0$ as $s_0 \rightarrow 0$ to $\xi \rightarrow 1$ as $s_1 \rightarrow 0$~\cite{suleimanov2011}.

\subsection{Machine-learning interatomic potential}

\subsubsection{Moment tensor potential}

MTP is the interatomic interaction model used as PES in this paper. It was described in detail in Refs. \cite{Shapeev2016-MTP,Gubaev2018-MTPR,gubaev2019-alloys}. Here we present only a brief description of MTP. 

We assume that our machine-learning interatomic potential is local, {\it i.e.}, the energy $E$ of each atomic configuration is partitioned into contributions $V$ of environments (neighborhoods) ${\bf \mathfrak{n}}_i$, $i=\overline{1,n}$ of each $i$-th atom: $E = \sum \limits_{i=1}^{n} V(\mathfrak{\bm n}_i) $. We expand each contribution through a set of basis functions: $V({\bf \mathfrak{n}}_i) = \sum \limits_{\alpha} \xi_{\alpha} B_{\alpha}({\mathfrak{\bm n}}_i)$, where $B_{\alpha}$ are the basis functions and $\xi_{\alpha}$ are the parameters we find after the training (fitting) of MTP (we describe the fitting in the end of this subsection). We construct the basis functions $B_{\alpha}$ as all possible contractions of the moment tensor descriptors yielding a scalar (see Ref.~\cite{gubaev2019-alloys} for details). The moment tensor descriptors have the following form
\begin{equation}\label{Moment}
M_{\mu,\nu}({\mathfrak{\bm n}}_i)=\sum_{j} f_{\mu}(|r_{ij}|,z_i,z_j) \underbrace {r_{ij}\otimes...\otimes r_{ij}}_\text{$\nu$ times},
\end{equation}
where ``$\otimes$'' denotes the outer product, $j$ enumerates all the atoms in the neighborhood $\mathfrak{\bm n}_i$, {\it i.e.}, within the distance less than $R_{\rm cut}$ from the $i$-th atom. Each neighborhood $\mathfrak{\bm n}_i$ is expressed by the interatomic vectors $r_{ij}$ and the types of $i$-th and $j$-th atoms: $z_i$ and $z_j$. The functions $f_{\mu}(|r_{ij}|,z_i,z_j)$ depend only on the neighborhood $\mathfrak{\bm n}_i$ and have the following form
\begin{align} \label{RadialFunction}
\displaystyle
f_{\mu}(|r_{ij}|,z_i,z_j) = \sum_{\beta} c^{(\beta)}_{\mu, z_i, z_j} T_\beta (|r_{ij}|) (R_{\rm cut} - |r_{ij}|)^2,
\end{align}
where $c^{(\beta)}_{\mu, z_i, z_j}$ is the one more set of MTP parameters to be fitted and $T_\beta (|r_{ij}|)$ are Chebyshev polynomials.

We denote the total set of parameters to be found by ${\bm \theta} := (\{ \xi_{\alpha} \}, \{ c^{(\beta)}_{\mu, z_i, z_j} \})$ and the MTP energy of a configuration ${\bm x}$ by $E = E({\bm \theta}; {\bm x})$. We find the parameters ${\bm \theta}$ by solving the following minimization problem
\begin{equation} \label{Fitting}
\begin{array}{c}
\displaystyle
\sum \limits_{k=1}^K \Bigl[ \left(E^{\rm AI}(\bm x^{(k)}) - E({\bm {\theta}}; \bm x^{(k)}) \right)^2 +
\\
\displaystyle
w_{\rm f} \sum_{i=1}^n \left| f^{\rm AI}_i(\bm x^{(k)}) - f_{i}({\bm {\theta}}; \bm x^{(k)}) \right|^2 \Bigr] \to \operatorname{min},
\end{array}
\end{equation} 
where $k$ enumerates all the configurations in the training set, $E^{\rm AI}$ and $f^{\rm AI}_i$ are the {\it ab initio} energy and forces, $w_{\rm f}$ is a non-negative weight which expresses the importantance of forces w.r.t. the energy in Eq.~\ref{Fitting}. We refer to the minimization problem Eq.~\ref{Fitting} as the fitting of MTP.

\subsubsection{Active learning}

In order to construct a global PES ({\it i.e.}, the PES which covers geometry regions relevant to the chemical process of interest), we should generate a training set that includes various representative configurations. In other words, we should decide whether a given configuration $\bm x^*$ generated during the RPMD trajectores is a candidate for adding to the training set ({\it i.e.}, whether this configuration is representative or not). To that end, we use the AL algorithm described below. 

Suppose we have $m$ parameters of MTP. Then we compose the following matrix
\[
\mathsf{B}=\left(\begin{matrix}
\frac{\partial E}{\partial \theta_1}\left( {\bm \theta}; \bm x^{(1)} \right) & \ldots & \frac{\partial E}{\partial \theta_m}\left({\bm \theta}; \bm x^{(1)}\right) \\
\vdots & \ddots & \vdots \\
\frac{\partial E}{\partial \theta_1}\left({\bm \theta}; \bm x^{(K)}\right) & \ldots & \frac{\partial E}{\partial \theta_m}\left({\bm \theta}; \bm x^{(K)}\right) \\
\end{matrix}\right),
\]
where each row in the training set corresponds to a particular configuration.

Next we select for training a subset of configurations yielding the most linearly independent rows in $\mathsf{B}$. This is equivalent to finding a square $m \times m$ submatrix $\mathsf{A}$ of the matrix $\mathsf{B}$ of maximum volume (maximal value of $|{\rm det(\mathsf{A})}|$). We do it using the so-called maxvol algorithm \cite{Zamarashkin}. In order to decide whether a given configuration $\bm x^*$ is representative or not, we calculate the extrapolation grade $\gamma(\bm x^*)$ defined as
\begin{equation} \label{Grade}
\begin{array}{c}
\displaystyle
\gamma(\bm x^*) = \max_{1 \leq j \leq m} (|c_j|), ~\rm{where}
\\
\displaystyle
c = \left( \dfrac{\partial E}{\partial \theta_1} (\bm \theta, \bm x^*) \ldots \dfrac{\partial E}{\partial \theta_m} (\bm \theta, \bm x^*) \right) \mathsf{A}^{-1}.
\end{array}
\end{equation}
This grade defines the maximal factor by which the above determinant can increase if ${\bm x^*}$ is added to the training set. Thus, if the configuration $\bm x^*$ is a candidate for adding to the training set then $\gamma(\bm x^*) \geq \gamma_{\rm th}$, where $\gamma_{\rm th} \geq 1$ is an adjustable threshold parameter which controls the value of permissible extrapolation. Otherwise, the configuration is not representative.

\subsection{RPMD-AL-MTP algorithm}

\begin{figure*} \begin{center}
\includegraphics[width=7.0in, height=4.0in, keepaspectratio=false]{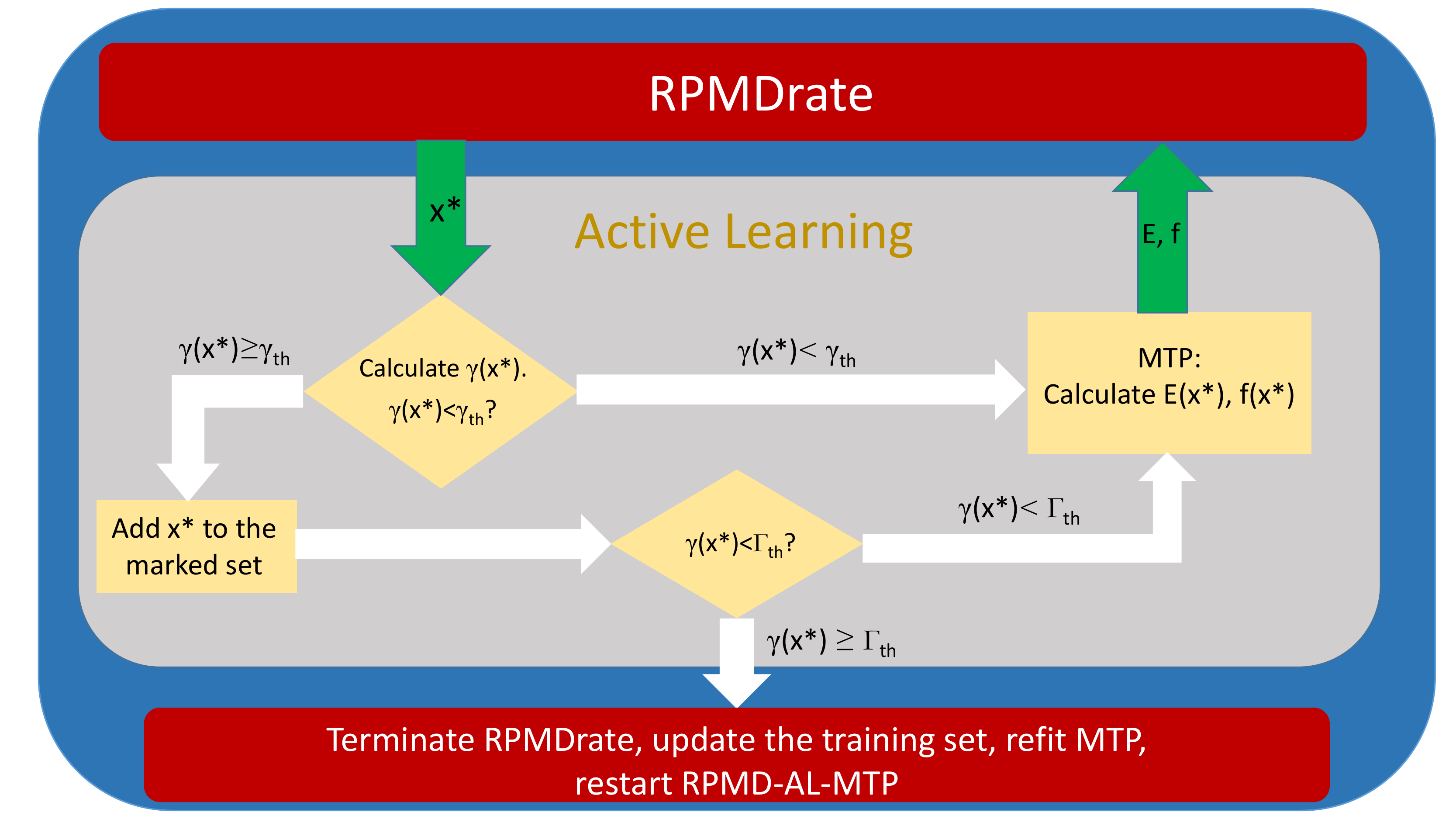}
\begin{center}
\caption{\label{Fig:figs0} RPMD-AL-MTP algorithm. For each configuration $\bm x^*$ occurring during the RPMD trajectory we calculate the extrapolation grade $\gamma(\bm x^*)$. If the extrapolation grade is small ($\gamma(\bm x^*) < \gamma_{\rm th}$) then this configuration will not be added to the training set, we simply continue the RPMD simulation. Otherwise, if the extrapolation grade is moderately high ($\gamma_{\rm th} \leq \gamma(\bm x^*) < \Gamma_{\rm th}$) then the configuration could be added to the training set and, thus, we mark this configuration and continue the RPMD run. Finally, if the extrapolation grade is too high, {\it{i.e.}}, $\gamma(\bm x^*) \geq \Gamma_{\rm th}$ then we terminate RPMDrate, update the training set with some of the marked configurations, refit the potential and restart the RPMDrate simulation.}
\end{center}
\end{center} \end{figure*}

Here we describe our combined RPMD-AL-MTP algorithm. We start by introducing two thresholds, namely, the lower bound $\gamma_{\rm th}$ and the upper bound $\Gamma_{\rm th}$ of permissible extrapolation, {\it i.e.}, $\gamma_{\rm th} < \Gamma_{\rm th}$.
The RPMD-AL-MTP algorithm continues as follows. For each configuration $\bm x^*$ occurring during an RPMD trajectory, we calculate $\gamma(\bm x^*)$. If $\gamma(\bm x^*) < \gamma_{\rm th}$ then $\bm x^*$ is not representative and therefore it will not be added to the training set. Hence, we just continue the RPMD simulation. Otherwise, this configuration could be added to the training set. If $\gamma_{\rm th} \leq \gamma(\bm x^*) < \Gamma_{\rm th}$ then $\gamma(\bm x^*)$ is sufficiently high for $\bm x^*$, but not too high to terminate the RPMD run. Hence, in this case, we mark the configuration $\bm x^*$ and add it in the marked set and continue the RPMD run. If $\gamma(\bm x^*) \geq \Gamma_{\rm th}$ then the extrapolation grade is too high, therefore, we terminate RPMD and add $\bm x^*$ to the marked set.
We then update the matrix $\mathsf{A}$ with the configurations from the marked set using the maxvol algorithm, calculate their {\it ab initio} energies and forces, add them to the training set, refit the potential, and repeat the entire RPMDrate simulation from the beginning (see Fig.~\ref{Fig:figs0}).

As a result, our algorithm will restart the RPMD simulations several times until the training set sufficiently covers the regions in the PES visited by RPMD trajectories during the simulation of chemical reaction.

\section{Application to S + H${_2}$}

\subsection{Computational details}

Here we describe the input parameters for the RPMD-AL-MTP algorithm. The RPMD simulations are performed using the RPMDrate code~\cite{suleimanov2013b}. 
The centroid PMF profiles were constructed along $\xi $ for the title reaction at 300, 400 and 500 K using the umbrella integration procedure~\cite{kastner05,kastner06}, that biases the dynamics simulation by dividing the reaction coordinate path into sampling windows. The Andersen thermostat~\cite{andersen80} was used in those trajectories. In order to calculate the ring polymer transmission coefficient, the recrossing trajectory evolution (with its centroid constrained at $\xi ^{\ddagger}$ that corresponds to the maximum free energy) was carried out using combination of parent-child trajectories and RATTLE algorithm~\cite{andersen83}. All input parameters of the RPMDrate simulation can be found in Tab.~\ref{tabs1}. We note that we took a smaller number of trajectories $N_{\rm trajectory}$ and fewer unconstrained (child) trajectories $N_{\rm totalchild}$ while running RPMD simulations with MTPs as compared to simulations with the original PES---our goal was to have very accurate reference results and make sure the MTP results converge within the typical accuracy of the RPMDrate computational procedure ($\leq 20 \%$). 
 
Since the main goal of the present study is to assess the applicability of the 
RPMD-AL-MTP algorithm, initially proposed for thermally activated chemical reactions,  to chemical reactions of insertion type, we have chosen one of the most typical representatives for benchmarking, namely, the S + H$_2$ system. 
We consider the PES of Ho et al.~\cite{sh2pes} used in the original RPMD study~\cite{rpmd_ins2} as the {\it ab initio} model for the present calculations and will refer to this model as the original PES. We emphasize that the rate coefficients calculated with this model were in a very good agreement with the experimental ones ~\cite{rpmd_ins2}.

Due to the reasons described below, we had to generate two MTPs. The first one contains 92 basis functions $B_{\alpha}$, 4 functions $f_{\mu}$ and 12 Chebyshev polynomials $T_{\beta}$. We denote this potential as MTP-286 (MTP with 286 parameters to be fitted). The second MTP is ``heavier'' than the first one as it includes 288 basis functions $B_{\alpha}$, 5 functions $f_{\mu}$ and 12 Chebyshev polynomials $T_{\beta}$. We denote it as MTP-530. As it could be seen from the Tab.~\ref{tabs1}, Fig.~{\ref{Fig:figs2}} and Fig.~{\ref{Fig:figs3}}, the accuracy of MTP-286 (for $T= 300$ K) was not high enough to detect the second dividing surface $s_1$ (the reaction coordinate $\xi ^{\ddagger}$) correctly due to rather small energy barrier ($\approx 10 ~\rm{meV}$) at the entrance to the complex-formation well which is typical for chemical reactions of insertion type.~\cite{rpmdreview} That is why we have fitted a ``heavier'' MTP and reached the accuracy needed to correctly detect the position of the free energy maximum (see the results in the next section). For both MTPs we took $R_{\rm cut} = 5 ~\angstrom$. The active learning was conducted with $\gamma_{\rm th} = 2$ and $\Gamma_{\rm th} = 10$, thus, we used the thresholds as in the original RPMD-AL-MTP work~\cite{novikov2018}.

As mentioned above, our aim is to compare the RPMD rate coefficients $k_{\rm RPMD}$ calculated using the original PES and the MTP PES. As described above, the calculations are dividied in two subsequent steps --- we first compute $k_{\operatorname{cd-TST}}$ at the free energy maximum and then $\kappa$ at $\xi ^{\ddagger}$. As in the previous study of thermally activated chemical reactions~\cite{novikov2018}, we generate two MTPs trained 
using two data sets from $k_{\operatorname{cd-TST}}$ and $\kappa$ calculations. 
Namely, in order to train the first MTP for calculating $k_{\operatorname{cd-TST}}$, we consider configurations from the reactant and complex formation regions ($\xi \in (-0.05, 1.05)$).
As a result, we obtain the first MTP that is very accurate for computing $k_{\operatorname{cd-TST}}$.

For calculating $\kappa$ and training the  second MTP, we take the training set obtained for the first MTP as a starting point and add configurations from the product region ($\xi > 1.05$).
Due to the insertion nature of the title reaction, many configurations from the product region were added to the training set from long time propagation of RPMD daughter trajectories (see Tab.~\ref{tabs2} and Fig.~\ref{Fig:figs1}) as opposed to the previously studied thermally activated reactions~\cite{novikov2018} which brings substantial difference between the two training sets.
As a result, the second MTP is less accurate if used for $k_{\operatorname{cd-TST}}$ due to excessive data from the product regions, however, is still sufficiently accurate for calculating $\kappa$ as the latter is not as sensitive to errors in the predicted energies and forces as $k_{\operatorname{cd-TST}}$ is.
The increased accuracy of the first MTP comes from the fact that it, essentially, interpolates the reference energies in a much smaller region of the multidimensional space.

After the calculations of $k_{\operatorname{cd-TST}}$ and $\kappa$, we obtain $k_{\rm RPMD}$ and compare the results obtained with the original PES and MTP PES.

\subsection{Transition state theory rates and transmission coefficients}

\begin{figure*} \begin{center}
\includegraphics[width=4.0in, height=5.0in, keepaspectratio=false]{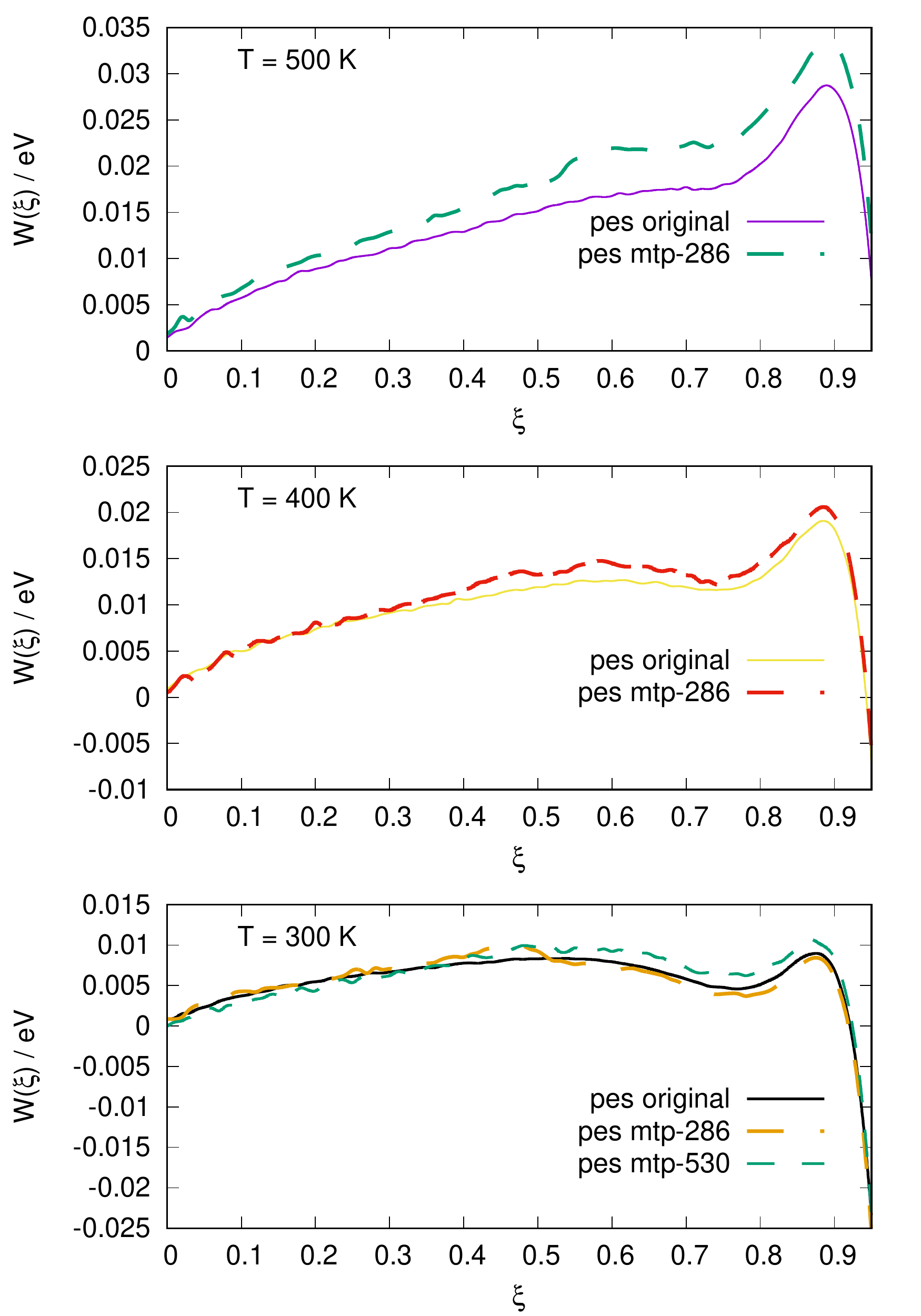}
\begin{center}
\caption{\label{Fig:figs2} Original (solid) and fitted (dashed) potentials of mean force for the S+H$_2$ reaction at $T = 300, 400, 500$ K. All the MTP profiles are close to the original ones.}
\end{center}
\end{center} \end{figure*}

\begin{figure*} \begin{center}
\includegraphics[width=4.5in, height=4.0in, keepaspectratio=false]{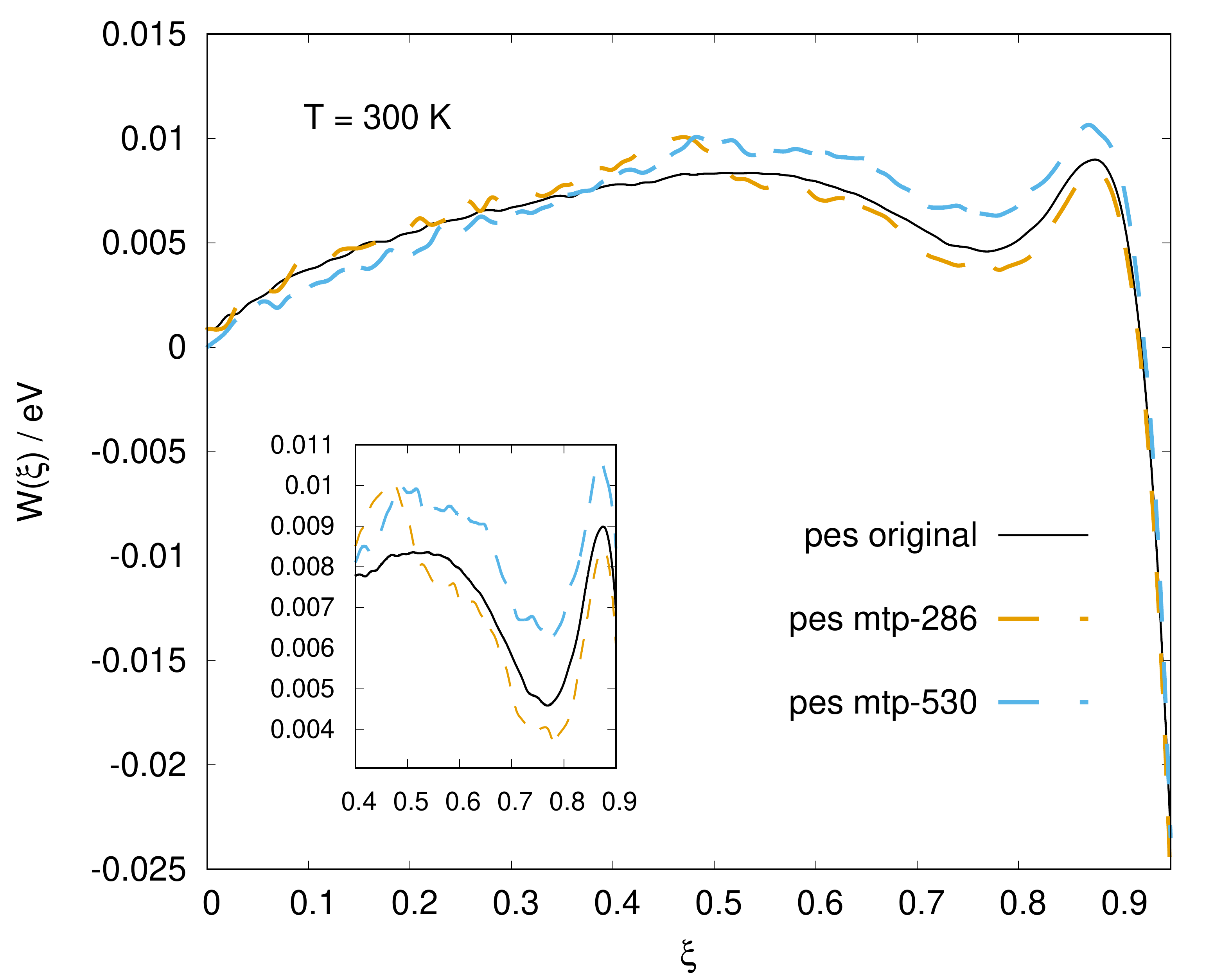}
\begin{center}
\caption{\label{Fig:figs3} Comparison between the original (solid) and two fitted (dashed) potentials of mean force for the S+H$_2$ reaction at $T = 300$ K. The two fitted PMFs obtained with MTP-286 and MTP-530. The ``heavier'' MTP (MTP-530) recognized the maximum of the free energy correctly.}
\end{center}
\end{center} \end{figure*}

\begin{figure*} \begin{center}
\includegraphics[width=5.5in, height=5.0in, keepaspectratio=false]{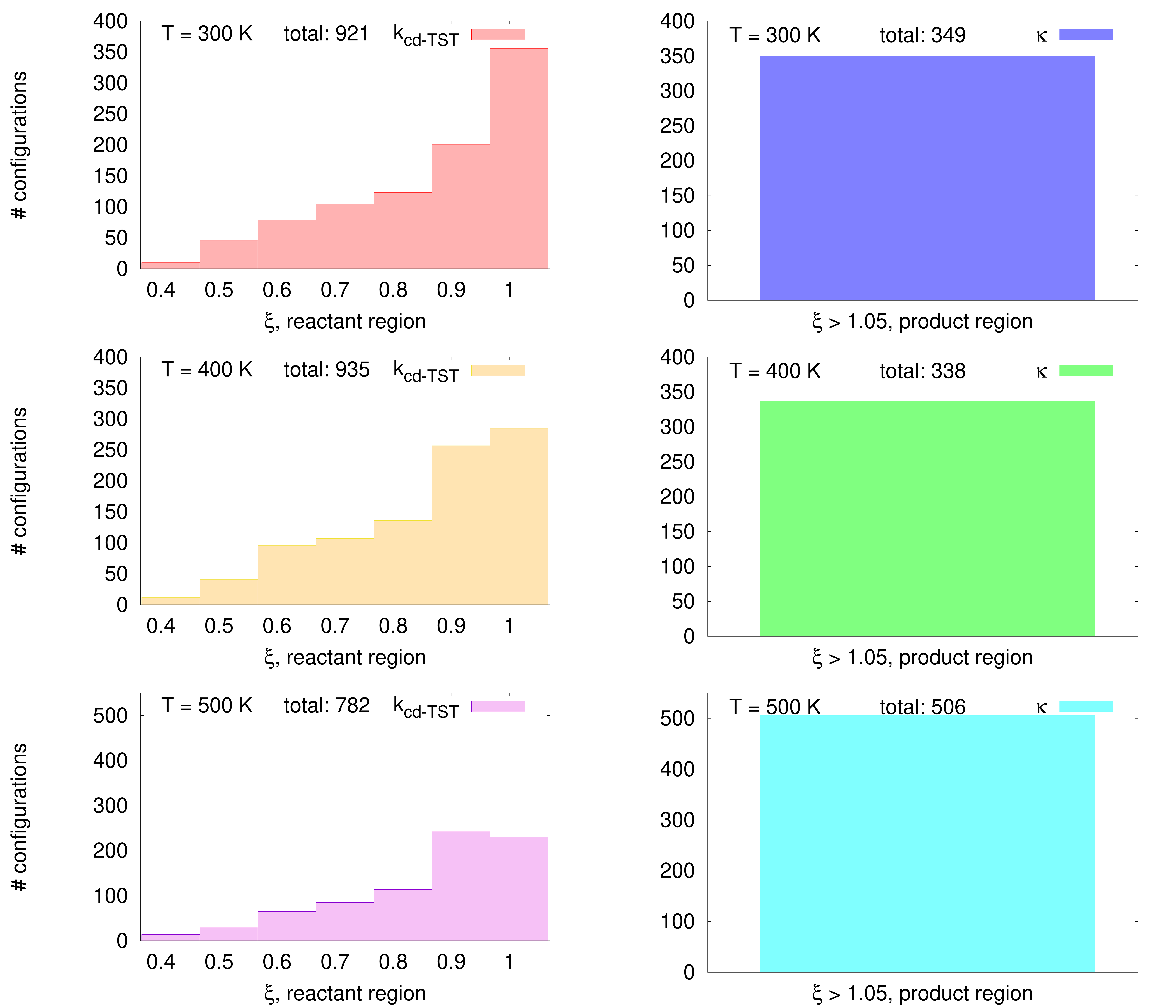}
\begin{center}
\caption{\label{Fig:figs1} The reactant and product set sizes for the S+H$_2$ system at $T = 300, 400, 500$ K. The numbers of configurations in the reactant region are given for the intervals (0.4, 0.5), (0.5, 0.6), \ldots , (0.9, 1.0), (1.0, 1.05), the total number of configurations in this region is shown at the top of the figure in the center. There was no selected configurations for $\xi < 0.4$.}
\end{center}
\end{center} \end{figure*}

The PMF profiles are shown in Figs.~\ref{Fig:figs2} and ~\ref{Fig:figs3}. For all the temperatures of interest, we can observe a rather smooth behavior of the MTP profiles which are close to the ones obtained with the original PES but with small deviations. Nevertheless, they do not affect the accuracy of the final output from these part of calculations --- cd-TST rate coefficients (the difference between the MTP and the original $k_{\operatorname{cd-TST}}$ is less than 10~$\%$, see Tab.~\ref{tabs3}). For $T = 300$ K, we found that the asymptotic barrier (located at $\xi \approx 0.47$) and the barrier near the complex formation (located at $\xi \approx 0.87$) ``compete'' with each other ({\it i.e.}, the values of $W(\xi)$ are close to each other near these points). This affects the calculation of $\xi ^{\ddagger}$ (see Tab.~\ref{tabs1}, Figs.~\ref{Fig:figs2} and ~\ref{Fig:figs3} for MTP-286, where 286 is the number of parameters in MTP) since the mother trajectory is constrained to $\xi ^{\ddagger}$~\cite{suleimanov2013b}. Thus, the initial configuration (at $\xi ^{\ddagger}$) for the second RPMD step obtained with MTP shifts to lower $\xi $ in comparison with the original PES. However, the final MTP and the original transmission coefficients are close to each other because the resulting RPMD rate coefficient does not depend on the choice of the dividing surface (see Ref.~\cite{suleimanov2011} and the discussion of the results on transmission coefficients below). In order to check whether we can increase the accuracy of the free energy profiles obtained using MTP and, therefore, to improve our estimation of $\xi ^{\ddagger}$, we decided to train a ``heavier'' MTP (MTP-530, see its description above). As a result, the free energy maximum shifted back to the original position observed previously (see Fig.~\ref{Fig:figs3}). 

\begin{figure*} \begin{center}
\includegraphics[width=4.0in, height=5.0in, keepaspectratio=false]{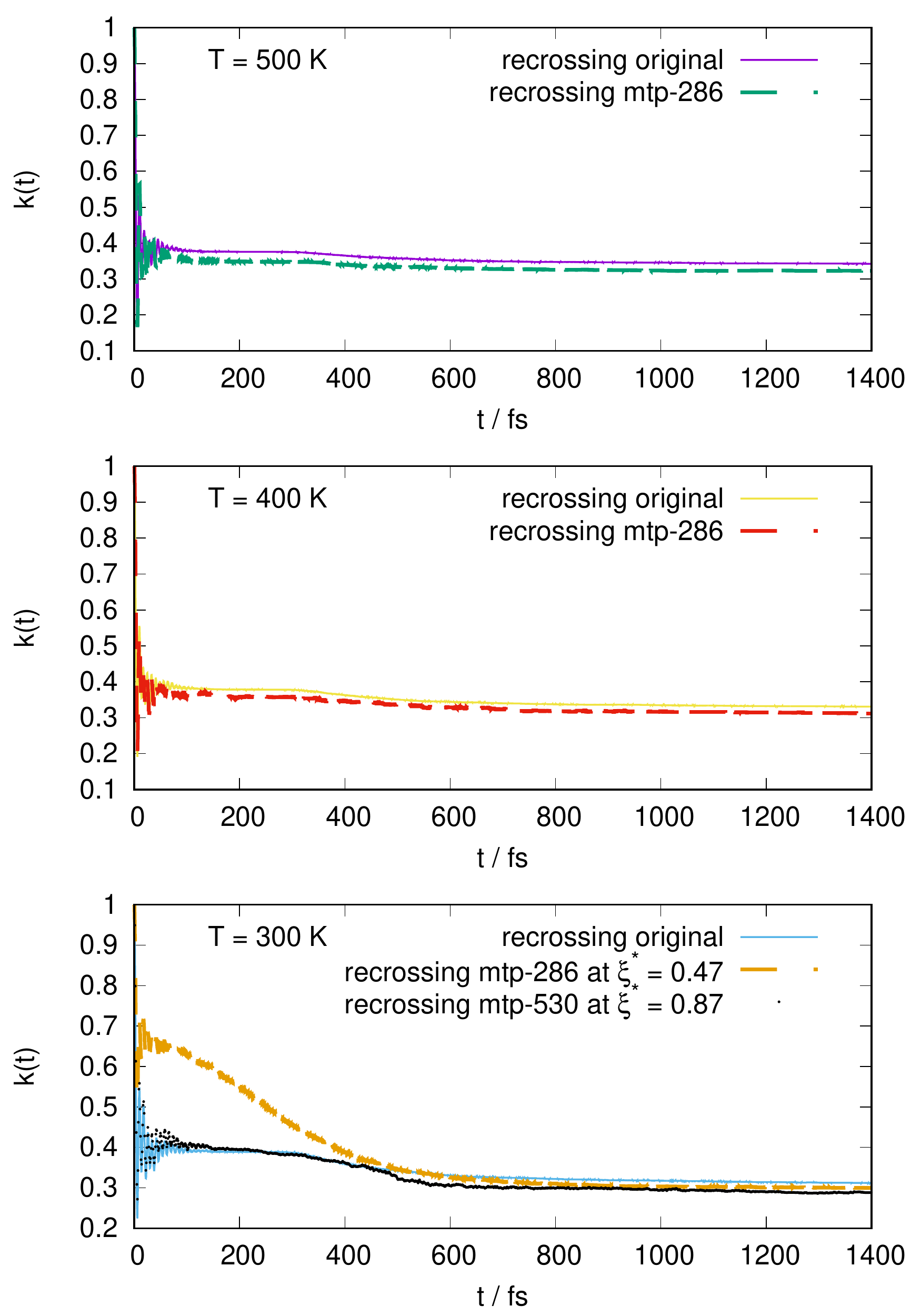}
\begin{center}
\caption{\label{Fig:figs4} Original (solid) and fitted (dashed) recrossing factors for the S+H$_2$ reaction at $T = 300, 400, 500$ K. The transmission coefficients obtained with MTPs and original potentials are close to each other.}
\end{center}
\end{center} \end{figure*}

The time-dependent TCs obtained with the MTPs and with the original PES are shown in Fig.~\ref{Fig:figs4}. 
The MTP and original TCs are in a very good agreement with each other --- the original time dependence is correctly reproduced by the present MTP calculations with only a small deviation of the plateau values, the difference between $\kappa^{\rm AI}$ and $\kappa^{\rm MTP}$ is less than 6 $\%$. We note that the profiles for the recrossing factors at $\xi ^{\ddagger} \approx 0.47$ (for MTP-286) and at $\xi ^{\ddagger} \approx 0.87$ (for MTP-530) differ from each other, nevertheless, the TCs obtained at $t \to \infty$ are close to each other, see Tab.~\ref{tabs3} and  Fig.~\ref{Fig:figs4}.

The resulting RPMD rate coefficients $k_{\rm RPMD}$ are summarized in Tab.~\ref{tabs3}. The difference between the original and present rate calculations is within the 6--14 $\%$ relative root-mean-square error. Apart from $T = 500$ K, the $k_{\operatorname{cd-TST}}$ and $\kappa$ calculations contribute equally to the total error, while at the highest temperature of the present study, the error in $k_{\operatorname{cd-TST}}$ increases. Nevertherless, the observed range of errors is comparable with the standard error of the RPMDrate computational procedure ($\leq 20 \%$)~\cite{suleimanov2013b}.
It is interesting to note that the accuracy of MTP-286 was better than that of MTP-530. This was because MTP-530 is three times more computationally expensive and therefore we took a smaller value of $N_{\rm trajectory}$ for MTP-530 than for MTP-286 to match the computational cost of the two potentials.

\section{Conclusions}
In the present paper we have shown that the combination of ring polymer molecular dynamics method and active learning of moment tensor potential (RPMD-AL-MTP) proposed and successfully tested on two representative thermally activated chemical reactions in [I. S. Novikov, Y. V.  Suleimanov, A. V. Shapeev, Phys. Chem. Chem. Phys. {\bf 20}, 29503-29512 (2018)] can also be applied to barrierless reactions. We have demonstrated that no significant changes were made to the RPMD-AL-MTP procedure, however, we found the following features. First, we need more time for training of MTP on the second RPMDrate step, namely, in the region of products. This is because  more time is needed for propagation of the trajectories as they go further in the product region compared to the ones in the case of thermally activated reactions. Moreover, we found that for complete reproduction of the original results at low temperatures it is necessary to use a ``heavy'' MTP ({\it i.e.}, trained on more data points with more parameters for fitting) in order to recognize the maxima of free energy profile correctly. Nevertheless, we note that the resulting rate coefficients obtained with the ``light'' and ``heavy'' MTPs do not differ significantly, since the result of the RPMD method does not depend on the choice of the dividing surface. 

In general, the relative deviation of the ring polymer rate coefficients obtained using the MTP PESs from those obtained using  the original PESs is within the range 6--14 $\%$. This error is comparable with the error obtained for thermally activated reactions, as well as with the typical error of the RPMDrate computational procedure ($\leq 20 \%$). 

To summarize, the present study completes our first attempt to combine RPMD with active learning of moment tensor potential [I. S. Novikov, Y. V.  Suleimanov, A. V. Shapeev, Phys. Chem. Chem. Phys. {\bf 20}, 29503-29512 (2018)] and demonstrates that the RPMD-AL-MTP method can be used for investigation of a gas-phase chemical reaction with any reaction path. In future, using our methodology, we plan to study more complex polyatomic chemical reactions. 

\section*{Author contributions}
Y.V.S. prepared the RPMDrate code for calculation of the considered chemical reaction rates and calculated the rates using original PES. I.S.N. and A.V.S. combined the RPMDrate code and the MLIP code and obtained the rate coefficients on MTP PES. All the authors discussed the results and wrote the paper.

\section*{Conflicts of interest}
The authors declare no conflicts of interest.

\begin{acknowledgments}
The work of I.S.N. and A.V.S. was supported by the Russian Science Foundation (grant number 18-13-00479). Y.V.S. thanks the European Regional Development Fund and the Republic of Cyprus for support through the Research Promotion Foundation (Projects: INFRASTRUCTURE/1216/0070 and Cy-Tera NEA ${\rm Y\Pi O\Delta OMH}$ / ${\rm \Sigma TPATH}$/0308/31). 
\end{acknowledgments}

\bibliography{rpmd_mtp_2019}

\begin{table*}
\scriptsize
\caption{\label{tabs1} Input parameters for the RPMD calculations on the S($^1$D) + H$_2$. 
The explanation of the format of the input file can be found in the RPMDrate code manual 
(\href{http://rpmdrate.cyi.ac.cy}{http://rpmdrate.cyi.ac.cy}).}
\begin{center}
\begin{tabular}{llll} 
\hline
\hline
Parameter & \multicolumn{2}{c}{Potential Energy Surfaces} & Explanation \\
\cline{2-4}
 & Original PES~\cite{sh2pes} & MTP-286$^a$/MTP-530$^c$  & \\
\hline
\multicolumn{4}{l}{Command line parameters} \\ 
\hline
\ttfamily{Temp}         & \multicolumn{2}{c}{300}      &       Temperature (K)  \\
                                 & \multicolumn{2}{c}{400}     &         \\
                                 & \multicolumn{2}{c}{500}       &         \\
\ttfamily{Nbeads}       & \multicolumn{2}{c}{128(300 K); 128(400 K), 128(500 K)} & Number of beads \\	
\hline
\multicolumn{4}{l}{Dividing surface parameters} \\ 
\hline
$R_\infty $  & 7.94 \AA   & 7.94 \AA   &  Dividing surface parameter (distance) \\
$N_{\rm bonds}$       & 1    & 1    & Number of forming and breaking bonds \\
$N_{\rm channel}$     & 2    & 2   & Number of equivalent product channels \\
S($^1$D)   & \multicolumn{2}{c}{(0.3757\AA,  2.1100\AA,  0.0000\AA )}  &  Cartesian coordinates (x, y, z) \\
H               & \multicolumn{2}{c}{(0.0000\AA,  0.0000\AA,  0.0000\AA )}  & of the intermediate geometry \\ 
H               & \multicolumn{2}{c}{(0.7514\AA,  0.0000\AA,  0.0000\AA )}  & \\ 
\hline
\ttfamily{Thermostat}  & 'Andersen' & 'Andersen'   &  Thermostat option \\
\hline
\multicolumn{4}{l}{Biased sampling parameters} \\ 
\hline
$N_{\rm windows}$  & 111    & 111        & Number of windows \\
$\xi _1$            & -0.05  & -0.05    & Center of the first window \\
$d\xi $              & 0.01   & 0.01      & Window spacing step      \\ 
$\xi _N$             & 1.05   & 1.05    & Center of the last window \\
$dt$  & 0.0001 & 0.0001   &  Time step (ps) \\
$k_i$  &  2.72  & 2.72  & Umbrella force constant ((T/K) eV) \\
$N_{\rm trajectory}$ & 80 & 15$^a$ (10$^c$)  & Number of trajectories \\ 
$t_{\rm equilibration}$ & 20  & 20    & Equilibration period (ps)  \\
$t_{\rm sampling}$ &  100 & 100    & Sampling period in each trajectory (ps) \\
$N_i$   & $2\times 10^8$ & $2\times 10^8$  & Total number of sampling points \\
\hline
\multicolumn{4}{l}{Potential of mean force calculation} \\ 
\hline
$\xi _0$            & 0.00  & 0.00   & Start of umbrella integration  \\
$\xi ^{\ddagger b}$             & 0.875 (300 K)  & 0.470 (300 K)$^a$ & End of umbrella integration \\
                              & & 0.872 (300 K)$^{c}$ \\
                                        & 0.885 (400 K) &  0.885 (400 K)$^{a}$ &   \\
                                        & 0.889 (500 K)  &  0.892 (500 K)$^{a}$ &    \\
$N_{\rm bins}$             & 5000  & 5000   &  Number of bins \\
\hline
\multicolumn{4}{l}{Recrossing factor calculation} \\ 
\hline
$dt$  & 0.0001 & 0.0001  &Time step (ps) \\
$t_{\rm equilibration}$  & 15  & 15   & Equilibration period (ps) in the constrained (parent)\\ 
 & &    & trajectory \\
$N_{\rm totalchild}$  & $10^5$  & ${10^4}$   & Total number of unconstrained (child) trajectories  \\
$t_{\rm childsampling}$  & 2  &  2  & Sampling increment along the parent trajectory  (ps)  \\
$N_{\rm child}$  & 100  & 100  &  Number of child trajectories per one   \\
  & & &  initially constrained configuration \\
$t_{\rm child}$  & 3  & 3   &Length of child trajectories (ps)  \\
\hline
\hline
\end{tabular}\\
\end{center}
$^a$ Obtained using MTP with 286 parameters (MTP-286). \\
$^b$ Detected automatically by RPMDrate. \\
$^c$ Obtained using MTP with 530 parameters (MTP-530). \\
\end{table*}

\begin{table*}
\caption{\label{tabs2} Number of configurations selected in the reactant region ($k_{\operatorname{cd-TST}}$ set size, the first training set), in the product region ($\kappa$ set size, the second training set), and the total training set size ($k_{\rm RPMD}$ set size) for the OH + H$_2$, CH$_4$ + CN \cite{novikov2018} systems and for the S + H$_2$ system. The ratio of configurations number in the product and the total training set is much greater for the barrierless reaction than for the thermally activated reactions.} 
\begin{center}
\begin{tabular}{c|c|c|c|c} \hline \hline
System, T, $n_{\rm beads}$ & $k_{\operatorname{cd-TST}}$ & $\kappa$ & $k_{\rm RPMD}$ & \multirow{2}{*}{$\dfrac{\kappa ~\rm{set ~size}}{k_{\rm RPMD} ~\rm{set ~size}}$ (\%)} \\ 
 & set size & set size & set size & \\ \hline
OH+H$_2$, 300 K, 128 & 1816 & 44 & 1860 & 2.4 \% \\ 
OH+H$_2$, 1000 K, 16 & 2014 & 83 & 2097 & 3.9 \% \\
CN+CH$_4$, 300 K, 128 & 4138 & 380 & 4518 & 8.4 \% \\ 
CN+CH$_4$, 600 K, 16 & 4572 & 320 & 4892 & 6.5 \% \\ \hline
S+H$_2$, 300 K, 128 & 921 & 349 & 1270 & 27.5 \% \\ 
S+H$_2$, 400 K, 128 & 935 & 338 & 1273 & 26.5 \% \\ 
S+H$_2$, 500 K, 128 & 782 & 506 & 1288 & 39.3 \% \\ \hline \hline
\end{tabular}
\end{center}
\end{table*}

\begin{table*}
\caption{\label{tabs3}Comparison of the centroid density transition state theory (cd-TST) rate coefficient $k_{\operatorname{cd-TST}}$, ring polymer transmission coefficient $\kappa$, and final rate coefficient $k_{\rm RPMD}$ calculated by the original PES (at $T = 300, 400, 500$ K), by MTP-286 (at $T = 300, 400, 500$ K) and by MTP-530 (at $T = 300$ K) for the S + H$_2$ system ($n_{\rm beads} = 128$). The accuracy of the resulting rate coefficient calculation is in the range from 6 $\%$ to 14 $\%$ which is comparable to the accuracy of the RPMD method. The accuracy of the resulting rate constant calculation obtained with MTP-530 ($T = 300$ K) is not as high as the accuracy obtained with MTP-286 because we took a smaller number of trajectories for MTP-530 due to its higher computational cost.} 
\begin{center}
\begin{tabular}{c|cccc} \hline \hline
 & \multicolumn{2}{c}{T = 300 K} & T = 400 K & T = 500 K  \\
 & MTP-286 & MTP-530 & MTP-286 & MTP-286 \\ \hline
 $k_{\operatorname{cd-TST}}^{\rm AI}$ (cm$^3$ s$^{-1}$) & \multicolumn{2}{c}{$2.64 \times 10^{-9}$} & $2.46 \times 10^{-9}$ & $2.48 \times 10^{-9}$  \\ 
$k_{\operatorname{cd-TST}}^{\rm MTP}$ (cm$^3$ s$^{-1}$) & $2.54 \times 10^{-9}$ & $2.47 \times 10^{-9}$ & $2.34 \times 10^{-9}$ & $2.25 \times 10^{-9}$  \\ 
error (\%) & 3.8 \% & 6.4 \% & 4.9 \% & 9.3 \% \\ \hline
$\kappa^{\rm AI}$ & \multicolumn{2}{c}{0.305} & 0.327 & 0.340  \\ 
$\kappa^{\rm MTP}$ & 0.297 & 0.287& 0.308 & 0.321 \\ 
error (\%) & 2.6 \% & 5.9 \% & 5.8 \% & 5.6 \% \\ \hline
$k_{\rm RPMD}^{\rm AI}$ (cm$^3$ s$^{-1}$) & \multicolumn{2}{c}{$8.05 \times 10^{-10}$} & $8.08 \times 10^{-10}$ & $8.44 \times 10^{-10}$  \\
$k_{\rm RPMD}^{\rm MTP}$ (cm$^3$ s$^{-1}$) & $7.54 \times 10^{-10}$ & $7.09\times 10^{-10}$ & $7.21 \times 10^{-10}$ & $7.22 \times 10^{-10}$  \\
error (\%) & 6.3 \% & 11.9 \% & 10.8 \% & 14.4 \% \\ 
\hline
\hline
\end{tabular}
\end{center}
\end{table*}

\end{document}